\documentclass{aa}  

\usepackage{graphicx}
\usepackage{txfonts}
\usepackage{lipsum}
\usepackage{subcaption}         
                                
\usepackage{lscape}             
                                
\usepackage{placeins}           
                                
\usepackage{hyperref}
\begin{document}

   \title{Star clusters as tracers of the perturbed southern \\ outer disk of the Large Magellanic Cloud}

   \author{Matías Chiarpotti\inst{1,2}, Andrés E. Piatti\inst{1,2}, Denis M.F. Illesca \inst{1,2} \and Roberto Butrón \inst{1} 
        }

   \institute{Instituto Interdisciplinario de Ciencias Básicas (ICB), CONICET-UNCuyo, Padre J. Contreras 1300, M5502JMA, Mendoza,
Argentina;
            \and Consejo Nacional de Investigaciones Científicas y Técnicas (CONICET), Godoy Cruz 2290, C1425FQB, Buenos Aires,
Argentina}

   \date{Received / Accepted}

  \abstract{The complex interaction history between the Large and Small Magellanic Clouds 
(L/SMC) has left significant imprints on their structural and chemical evolution. In this 
work, we use a sample of 70 star clusters located in the southern outer disk of the LMC 
($R > 5^{\circ}$) to trace the magnitude of these tidal perturbations. Using deep $g,i$ 
photometry from the SMASH DR2 survey and the Automated Stellar Cluster Analysis \texttt{(ASTeCA)}
 code, we provide a homogeneous determination of cluster ages, metallicities, and 
heliocentric distances. Our results show a population with ages ranging from 1.02 to 3.34 
Gyr and metallicities between -0.70 and -0.07 dex. We find that clusters at greater 
heliocentric distances are predominantly located toward the western regions, facing the 
SMC. Analysis of the vertical distribution reveals an oscillatory variation in the mean 
height ($\langle Z \rangle$) relative to the LMC mid-plane and an increase in intrinsic 
vertical dispersion ($W$) toward the West. These geometric features are consistent with a
 corrugated disk structure induced by recent tidal encounters with the SMC. Chemically, 
the clusters follow the age-metallicity relationship typical of the LMC and are 
well-represented by a bursting-type formation model. The lack of chemically distinct 
clusters suggests that while the interaction was sufficient to dynamically deform and 
corrugate the outer disk, it did not lead to significant cluster capture from the SMC.
 We conclude that the southern outer disk remains structurally and chemically part of
 the LMC, despite being heavily perturbed by tidal forces.
 }

   \keywords{techniques: photometric --  galaxies: individual: LMC -- galaxies: star clusters }

\titlerunning{LMC star clusters}

\authorrunning{M. Chiarpotti et al.}

\maketitle

\markboth{M. Chiarpotti et al.: }{LMC star clusters}

\section{Introduction}

The Magellanic Clouds is a pair of dwarf galaxies with a complex
interaction history \citep{choietal2022,munozetal2023}. Their
proximity to the Milky Way has allowed us to resolve individual stars
and hence has favored targeting them as valuable astrophysical
laboratories to study galaxy interactions. Indeed, since recent years 
the aftermaths of the successive tidal encounters between the Large and 
the Small Magellanic Clouds (L/SMC) have caught our attention 
\citep[e.g.][]{cullinane2022magellanic}. As far as the LMC is concerned, several 
stellar substructures have been discovered in the outer galaxy 
regions with claimed tidal origins. \citet{mackey2018substructures} 
showed that the outer LMC disk is strongly distorted, with irregular shapes 
which are evidence of warping, and that the western side facing the SMC
exhibits an important truncation. They also found low surface-brightness 
stellar substructures toward the northern and southern LMC disk periphery. 
\citet{choi2018smashing} reinforced these findings using the Survey of the 
Magellanic Stellar History \citep[SMASH,][]{nidever2021second} Data Release 
2, and concluded that the LMC disk is more notably warped 
from $\sim$ 5$\degr$ from the LMC center.

Kinematic studies based on proper motions and/or radial velocities of
individual stars distributed throughout the LMC disk, and particularly across 
the recently discovered stellar substructures, show that the latter have been 
originated due to consecutive interactions with the SMC \citep{cullinane2022magellanic,cullinane2022magellanicb,schmidt2022vmc}.
These outcomes have significantly changed our knowledge about the formation
and dynamical evolution of the LMC. However, the new picture of the LMC relies 
basically on observations of some galactic stellar populations, namely:
red clump stars, red giant branch stars, Carbon stars, among others.
As far as we are aware, star clusters have not been used to trace the
internal kinematics of the LMC, with the exception of \citet{piatti2019two}, who
concluded that the 15 LMC ancient globular clusters pertain to two kinematically
different populations. Star clusters present in general some advantages over field 
stars, whose selection criteria can lead to samples contaminated by Milky Way 
field stars \citep{jimenezarransetal2023}. For instance, their radial velocities 
and mean proper motions are derived from the average of several measurements of 
cluster members, resulting in statistically sound mean values. Star clusters are 
also appropriate proxies for the internal motions of the LMC, compared to the 
radial velocities and mean proper motions of stars located along the line of sight 
across the entire extent of the galaxy. Furthermore, star clusters have precisely 
determined ages and metallicities, so the internal kinematics of the LMC can be 
easily linked to the ages and metallicities of the star clusters, thus providing 
a suitable framework for our understanding of the galaxy's chemical and kinematic 
formation and evolution.

\begin{figure}
    \centering
    \includegraphics[width=\columnwidth]{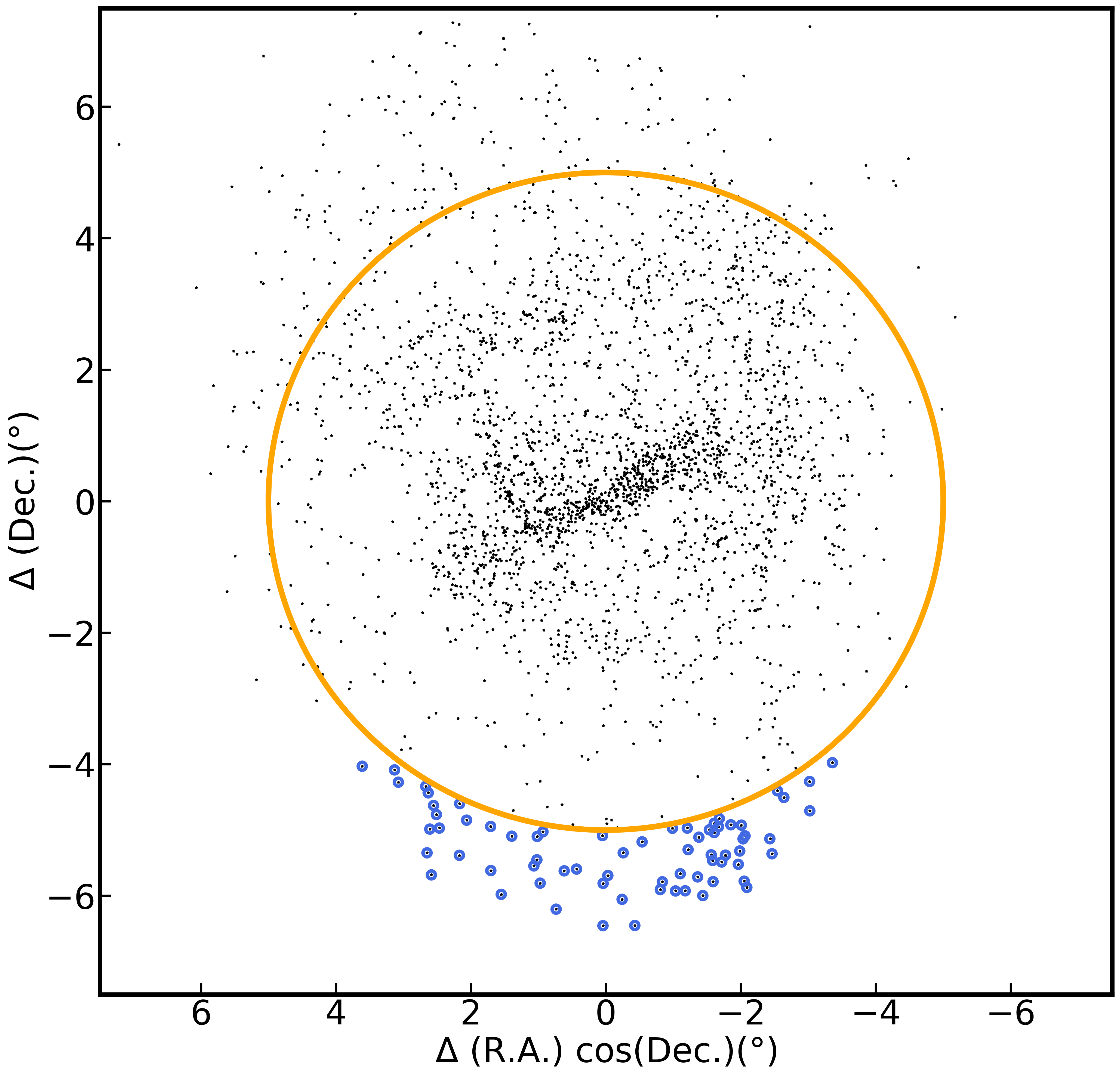}
    \caption{Sky distribution of star clusters of the LMC \citep{bica2008general} represented 
by black dots. Large open blue circles represent the star clusters studied in this work, while
 the orange circle corresponds to a radius of 5$\degr$ from the LMC center. North is up, and 
East is to the left.}
    \label{fig1}
\end{figure}

Among the LMC outer disk regions of particular interest for studying the 
magnitude and intensity of gravitational perturbations caused by tidal 
interactions with the SMC, the disk region located at radii larger than 5$\degr$ from 
the LMC center stands out. It spans position angles between $\sim$ 130$\degr$ and 
230$\degr$, measured from North to East. This is because the 
Magellanic Bridge, which connects the two Magellanic Clouds and is a privileged 
witness to the mutual interaction between these galaxies, is embedded in that
part the LMC \citep{mackey2018substructures,massana2024magellanic}. Therefore, the 
star clusters that populate it (see Figure~\ref{fig1}) are ideal objects for 
characterizing their mutual interaction. Precisely, in this work we homogeneously
estimate ages, metallicities and heliocentric distances of almost all star clusters
cataloged by \citet{bica2008general} located in that region of the LMC southern 
periphery, with the aim of connecting their value distributions with the Magellanic 
Clouds interaction history.

Section 2 describes the observational data used in this work and its treatment, detailing the 
photometric processing from SMASH DR2 and the procedure for the statistical 
decontamination of field stars. Section 3 addresses the determination of the clusters' 
fundamental astrophysical parameters, namely: age, metallicity, and heliocentric 
distance, through the use of the  Automated Stellar Cluster Analysis code 
\citep[\texttt{ASTeCA},][]{perren2015asteca}, as well as the validation of 
these results by comparing them with values from the literature. Section 4 presents 
the analysis and discussion of the three-dimensional spatial distribution of the 
clusters and their age-metallicity relationship (AMR). Finally, Section 5 summarizes 
the main conclusions obtained in this work.

\section{Data handling}

This section describes in detail the acquisition and subsequent processing of the photometric 
information for each cluster, aimed at deriving their corresponding astrophysical parameters. 
Inside the selected area described above, \citet{bica2008general} reported
a total of 71 star clusters (see open blue circles in 
Figure~\ref{fig1}). Out of these, we selected 70 clusters with photometric data in SMASH DR2,
SL~36 being the exception. We found astrophysical parameter estimates for 21 clusters, which
are listed in Table~\ref{Table1}, along with their respective references.

\begin{figure}
    \centering
    \includegraphics[width=\columnwidth]{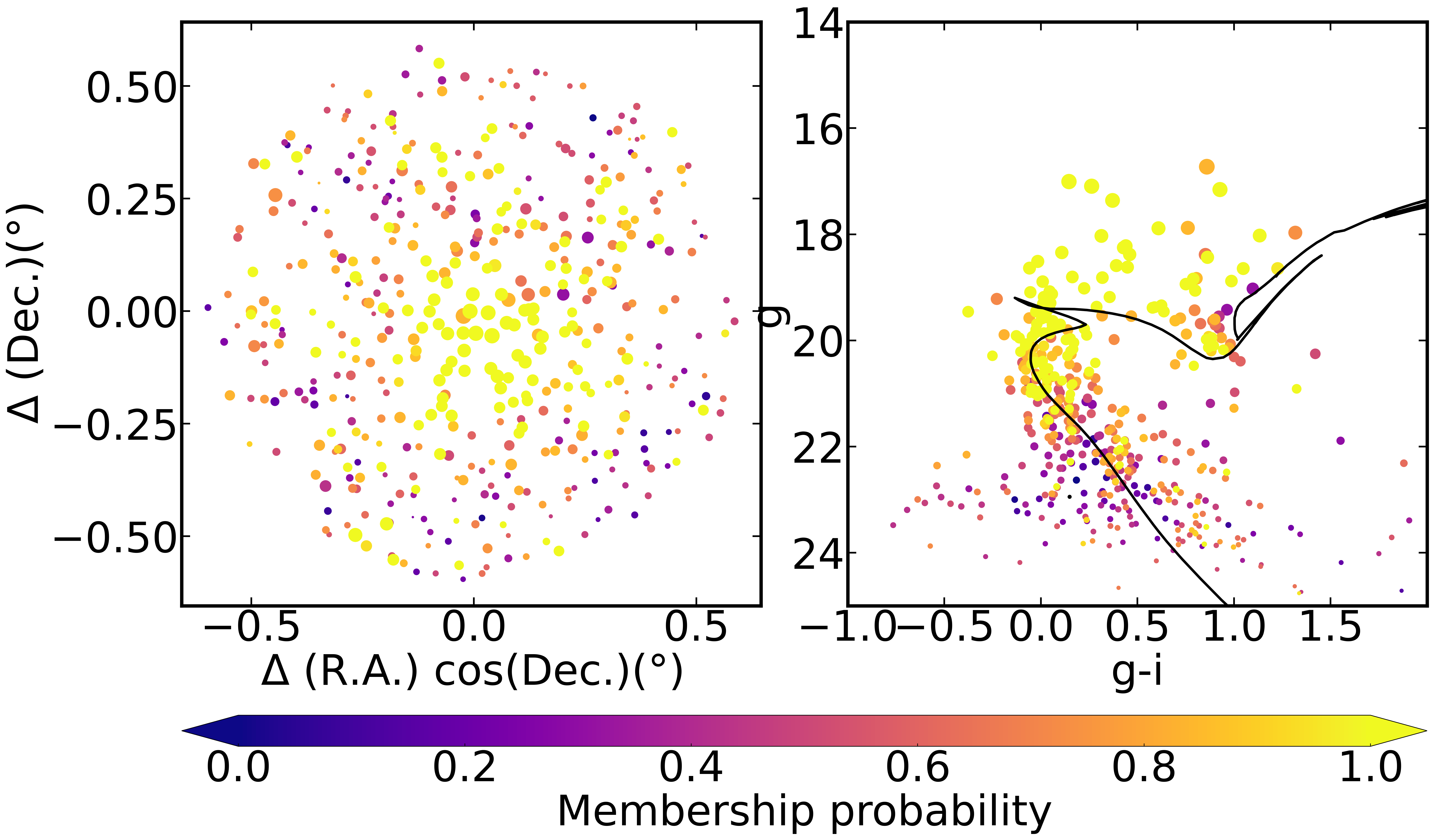}
    \caption{{\it Left:} Sky distribution of stars in the field of IC~2134. The symbol size is 
proportional to the brightness of the star. {\it Right:} Observed cluster CMD with the isochrone 
\citep{bressan2012parsec} corresponding to the derived fundamental parameters superimposed (see
 details in Sect.~2). Symbol colors refer to the membership probability assigned to each star.}
    \label{fig2}
\end{figure}

We employed the SMASH DR2 database to retrieve the necessary information. This survey utilizes 
the Dark Energy Camera (DECam) on the Victor Blanco (4m) telescope at Cerro Tololo Interamerical 
observatory (CTIO, Chile), to survey the Magellanic Clouds, reaching depths of approximately 
magnitude 24 in the $ugriz$ filters, with the goal of identifying low-surface-density stellar 
populations widely distributed across the galaxies. Specifically, we used the magnitudes in 
the $g$ and $i$ filters, which reach depths of 24.8 and 24.2 mag, respectively, being appropriate 
for our analysis.
To download the data, we used the Astro Data Lab interface, part of the Community Science and 
Data Center at NSF’s National Optical-Infrared Astronomy Research Laboratory. From this platform, 
we extracted the coordinates (R.A., Dec.), the magnitudes in the $g$ and $i$ filters
 along with their respective errors, the interstellar reddening $E(B-V)$, the $\chi$ and 
$\textsc{sharpness}$ parameters. 
These parameters were extracted for every star located within a radius six times larger than that 
reported by \citet{bica2008general} for these star clusters, with the aim of obtaining a 
representative sample of the surrounding field stars. Furthermore, following the criteria 
described by \citet{nideveretal2017}, we selected sources satisfying $0.2 \leq \textsc{sharpness} 
\leq 1.0$ and $\chi < 0.5$. The \textsc{sharpness} criterion was used to exclude extended sources, 
while $\chi$ was used to remove sources whose stellar profiles show significant deviations from the 
best-fitting model.

\begin{figure}
    \centering
    \includegraphics[width=0.8\columnwidth]{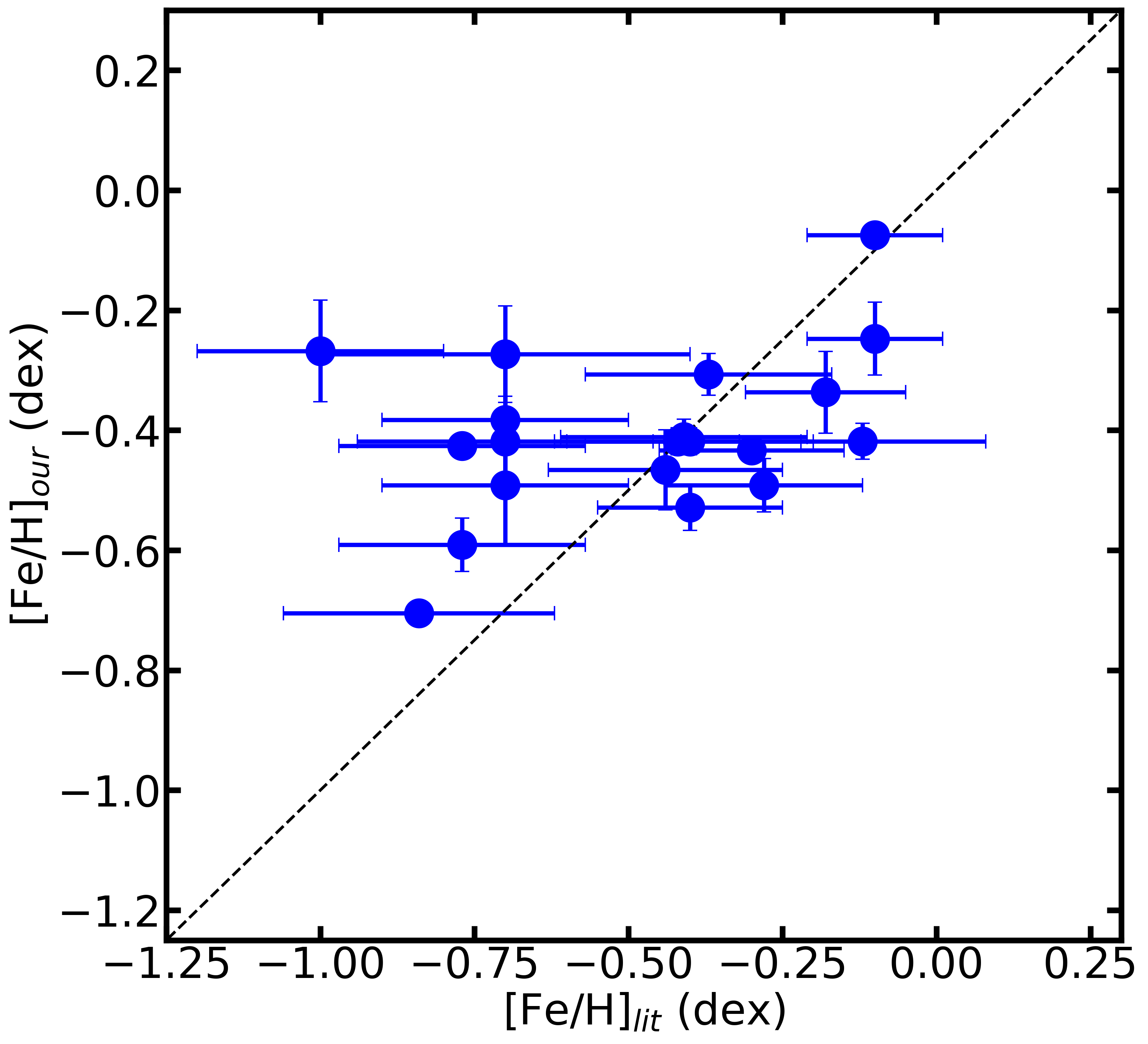}
    \caption{Comparison between literature ([Fe/H]$_{lit}$) and derived ([Fe/H]$_{our}$)
    metallicities. Error bars represent the uncertainties associated with the corresponding metallicity 
    determinations. The identity relationship is represented by a dashed line.}
    \label{fig3}
\end{figure}

We proceeded to remove the stars belonging to 
the field with the aim of obtaining a decontaminated color-magnitude diagram (CMD)
for each cluster. The process used was 
designed by \citet{piatti2012washington}, where the required input are the reddening-corrected 
$g_0$ and $i_0$ magnitudes, calculated from the observed $g$ and $i$ magnitudes, the $E(B-V)$ 
values obtained from the SMASH survey, and the $A_{\lambda}/E(B-V)$ ratios for $\lambda = g, i$ 
provided by \citet{abott2018dark}.
The process began with the selection of stars located within the clusters' radii
\citep{bica2008general}. Subsequently, 
1000 CMDs were generated corresponding to 1000 stellar fields randomly selected at a distance of 
2.5 radii from the cluster center. Each reference field CMD was built with all the measured
stars within circles of radius equal to that of the respective star cluster.
Next, the cleaning of a cluster CMD was performed by removing 
stars whose magnitude and color are the closest ones to those in the reference field CMDs.
This procedure was repeated for each of the 1000 reference field CMDs, thus obtaining
1000 different cleaned cluster CMDs. Finally, each star was assigned a membership probability  
based on the number of times it was removed during the cleaning process 
(see the color bar in Figure~\ref{fig2}). This probability was defined as P = 1 - N / 1000, where 
N represents the number of times a star was subtracted during the cleaning process. Stars with 
membership probabilities P > 0.5 were considered probable cluster members in the
subsequent analysis.

\section{Astrophysical properties}

Once the decontamination process for each cluster was completed, we derived their astrophysical
 parameters using routines of \texttt{ASTeCA}. It estimates parameters such as metallicity, age,
 distance modulus, and reddening, among others, by fitting  synthetic CMDs
that it generates. The astrophysical parameters associated to the best fitted synthetic CMD are 
taken as the optimal properties of the star cluster. To obtain the synthetic CMDs, we used the 
initial mass function proposed by \citet{kroupa2002initial} and fixed a minimum mass
 ratio for binary systems of 0.5. The simulations covered star cluster masses between 100 
and 5000~$M_\odot$. To simulate the stellar populations, 
we employed PARSEC v1.2S isochrones \citep{bressan2012parsec} adapted to the SMASH photometric 
system. These cover ages from log(age/yr) = 6.6 to 10.1, with increments of
 $\Delta$log(age/yr) = 0.025, and metallicities from Z = 0.0001 ([Fe/H] = -2.18 dex) to Z = 0.030
 ([Fe/H] = 0.30 dex) in steps of $\Delta$Z = 0.001. This range of parameters covers the entire 
known age-metallicity relationship of the LMC \citep{Piatti2013}. In some cases, these ranges 
were restricted to obtain more accurate results. The uncertainties associated with the derived
astrophysical parameters were estimated using the standard bootstrap method implemented
within  \texttt{ASTeCA}. Figure~\ref{fig2} illustrates,
in the left panel, the sky distribution of stars in the field of the star cluster IC~2134. The 
right panel presents its observed CMD, with the theoretical isochrone corresponding to the
best fitted astrophysical properties. The color bar indicates the membership probability of each 
star. The resulting clusters' parameters are listed in Table~\ref{Table2}. 

\begin{figure}
    \centering
    \includegraphics[width=0.8\columnwidth]{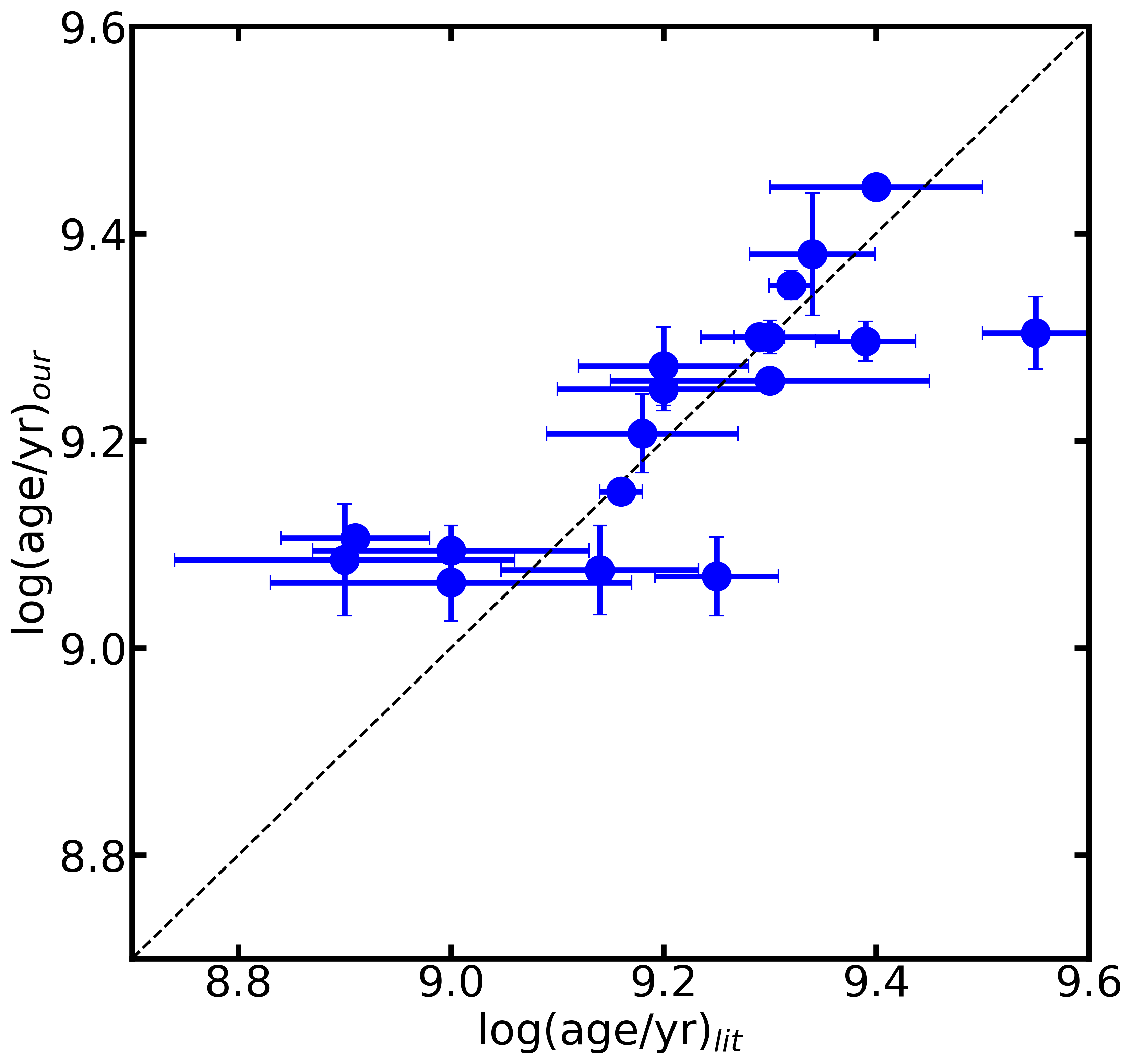}
    \caption{Comparison between literature (log(age/yr)$_{lit}$) and derived (log(age/yr)$_{our}$)
    ages. Error bars represent the uncertainties associated with the corresponding metallicity 
    determinations. The identity relationship is represented by a dashed line.}
    \label{fig4}
\end{figure}

Figures~\ref{fig3} and~\ref{fig4} illustrate the comparison between the metallicities
and ages obtained in this work and those reported in the literature, respectively, with error
bars included for both datasets. Although some differences are found between our determinations
and the literature values, our uncertainties are smaller in all cases, allowing more precise
estimates of the astrophysical parameters.
We derive a mean metallicity difference of $\Delta$[Fe/H] = |[Fe/H]$_{our}$ - [Fe/H]$_{lit}$| = 0.18
$\pm$ 0.15 dex, with a Pearson correlation coefficient of $r$ = 0.38. For
the ages, we obtain $\Delta$log(age/yr) = |log(age/yr)$_{our}$-(log(age/yr)$_{lit}$)| = 0.045
$\pm$ 0.028, with a Pearson correlation coefficient of $r$ = 0.78. Regarding heliocentric distances
(see Table~\ref{Table1}), we obtained $\Delta$$(m-M)_0$ = |$(m-M)_0$$_{our}$-$(m-M)_0$$_{lit}$| =
0.12$\pm$ 0.08 mag.

\section{Analysis and discussion}

Table~\ref{Table2} summarizes the calculated astrophysical parameters (heliocentric 
distance, metallicity and age) for the 70 clusters analyzed, along with their respective
 uncertainties. The second column includes the reddening values, which were extracted directly 
 from the SMASH database.

\begin{figure}
    \centering
    \includegraphics[width=\columnwidth]{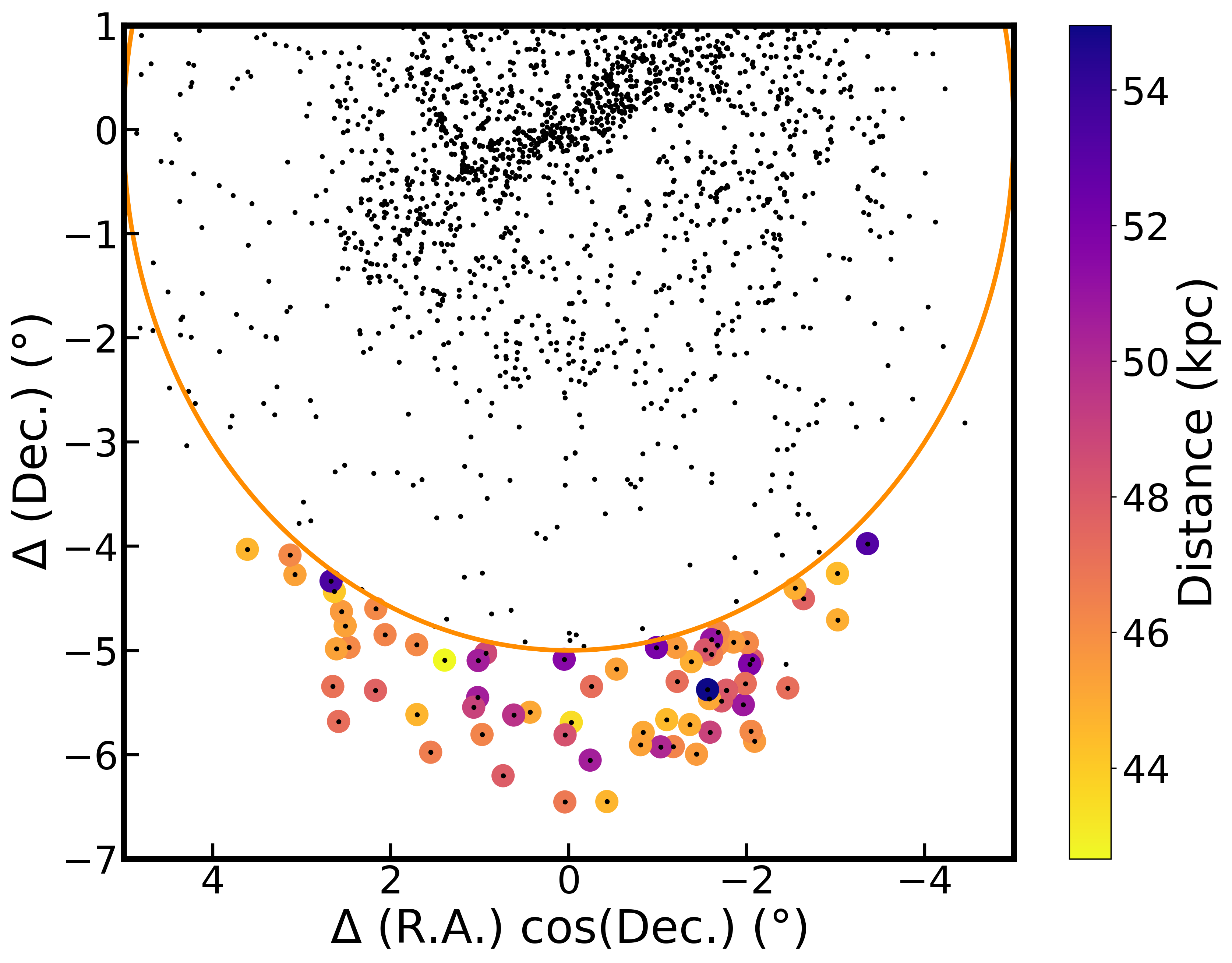}
    \caption{Spatial distribution of LMC star clusters \citep{bica2008general}, represented by 
black dots. Colored points indicate the heliocentric distances of the star clusters derived 
in this work (see Table~\ref{Table2}), with the color bar showing the distances in kpc. 
The orange circle corresponds to a radius of 5$\degr$ from the LMC center. North is
 up, and East is to the left.}
    \label{fig5}
\end{figure}

Our results for the cluster ages show a range between 1.02 and 3.34 Gyr, with an average value 
of 1.72 Gyr. Metallicities fall within the interval of -0.70 to -0.07 dex, with a 
mean value of -0.40 dex. Heliocentric distances vary from 42.66 kpc for the nearest 
cluster to 54.95 kpc for the most distant, as shown in Figure~\ref{fig5}, with an average 
distance of 47.16 kpc. 
The right panel of Figure~\ref{fig2} illustrates the best isochrone fit obtained for IC~2134. We 
derived a metallicity of [Fe/H] = -0.26 $\pm$ 0.04 dex, an age of log(age/yr) 
= 9.09 $\pm$ 0.02, and a heliocentric distance of 43.45 $\pm$ 1.20 kpc. This same fitting
procedure was applied to the remaining clusters in the sample, and the corresponding figures 
are available at \href{https://doi.org/10.5281/zenodo.22776595}{online supplementary 
material}.

\begin{figure} 
    \centering
    \includegraphics[width=\columnwidth]{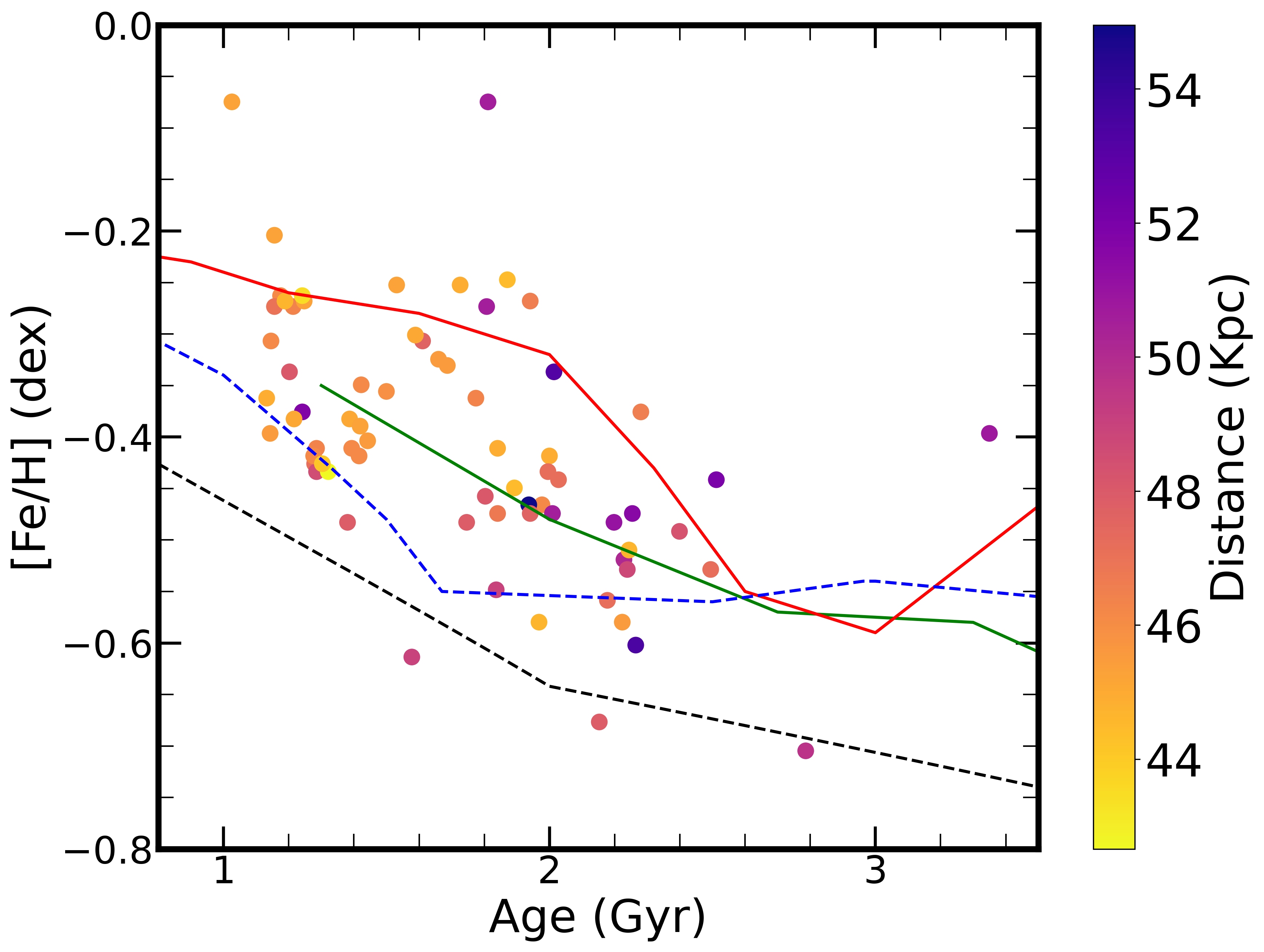}
    \caption{Age-metallicity relationship of the star clusters analyzed in the southern 
region of the LMC outer disk. The color bar indicates the heliocentric distances of the 
clusters. Solid lines represent previous observational relationships: the solid green line 
corresponds to the observational AMR for field stars from \citet{Piatti2013}, and the 
solid red line represents the relationship derived exclusively from star clusters by 
\citet{perren2017}. Dotted lines indicate theoretical chemical evolution models: the blue 
line represents the bursting model from \citet{pagel1998chemical}, and the black 
line denotes a closed-box evolutionary model calculated by \citet{geha1998stellar}.}
    \label{fig6}
\end{figure}

Figure~\ref{fig5} presents the spatial distribution of the 70 analyzed star clusters, 
color-coded by 
their heliocentric distances. A trend is observed where clusters at greater distances 
are predominantly located toward the western regions. 
Figure~\ref{fig6} shows the AMR diagram constructed from our results, 
where the color bar indicates the heliocentric distances associated with each cluster. 
We find that younger clusters exhibit higher metallicities and tend to be located 
at smaller heliocentric distances. 
We overlaid four curves representative of the LMC AMR onto Figure~\ref{fig6}, divided into two groups: 
solid lines correspond to observationally derived relationships, while dotted lines represent 
theoretical models. The solid green curve shows the AMR obtained from field stars by 
\citet{Piatti2013}. The solid red line represents the AMR 
derived by \citep{perren2017}, constructed exclusively from a sample of star clusters. 
Regarding the theoretical models, the blue dotted line represents the chemical evolution of 
the LMC calculated by \citet{pagel1998chemical} under a bursting scenario. 
Conversely, the black dotted line corresponds to the evolutionary model from 
\citet{geha1998stellar}, based on a closed-box scenario. As can be seen, the observational 
AMRs and that of the bursting model satisfactorily reproduce the distribution obtained 
in this work; the closed-box model shows more significant discrepancies. \\

In the following, we analyze the results obtained for the studied clusters, with a particular 
emphasis on their spatial distribution perpendicular to the LMC plane, 
as well as their AMR. These results are compared with 
previous studies to evaluate how the LMC 
structure and chemical properties may have been affected by recent interactions with the 
SMC \citep[e.g.][]{besla2012role,olsen2011population}.
Unlike previous studies based primarily on field star populations, such as red clump or red 
giant branch stars \citep{van2001magellanic,choi2018smashing}, this work utilizes star clusters as tracers 
of the outer disk structure. Because clusters possess well-determined ages, metallicities, 
and distances, they allow us for a reliable connection between the dynamical and chemical 
evolution of the LMC outer disk.

\subsection{Spatial analysis}

To analyze the vertical structure of the studied clusters to the LMC disk, we transformed the obtained
 heliocentric distances into heights relative to the LMC plane ($Z$) using the 
equation proposed by \citet{van2001magellanic}. We adopted a distance to the galaxy center 
of $D_0$ = 49.59 kpc \citep{pietrzynski2019distance} and the geometric parameters determined by 
\citet{van2014third}, namely: inclination $i$ = (34.0 $\pm$ 7.0)$\degr$, position angle 
$\theta$ = (139.1 $\pm$ 4.1)$\degr$, and center coordinates R.A. = 79.88$\degr$ and Dec. 
= -69.59$\degr$, respectively. These parameters were chosen because they were obtained from a 
large sample of intermediate- and old-age field stars, including carbon, AGB, and RGB stars, 
distributed across the entire LMC field. Consequently, they provide a robust estimate of the mean 
symmetry plane of the galaxy, making them particularly suitable for comparison with the cluster 
population analyzed in this work. 
We tested the sensitivity of the derived Z values to the uncertainties in the adopted parameters 
and found that they do not affect the overall vertical distribution of the 
clusters. Most of the variation in the derived Z values is associated with the 
uncertainties in the heliocentric distances, which were taken into account through Monte Carlo 
simulations.
We adopt a coordinate system in which positive $Z$ values indicate positions below the LMC 
mid-plane, 
while negative values correspond to positions above it. In this framework, the larger
the $Z$ values, the closer the clusters to us.

To characterize the vertical distribution of the clusters, we employed a maximum 
likelihood method 
\citep{walker2006internal,pryor93} to estimate the mean value $\langle Z \rangle$ and the 
intrinsic
 dispersion $W$, using the individual uncertainties $\sigma_Z$ propagated over the value 
of $Z$ via the following expression:\\

$L = \prod_{i=1}^{N} [2\pi(\sigma_{Z,i}^2 + W^2)]^{-1/2} \exp \left[ -\frac{(Z_i - \langle Z \rangle)^2}{2(\sigma_{Z,i}^2 + W^2)} \right]$,\\

\noindent where $Z_i$ and $\sigma_{Z,i}$ correspond to the height and its 
uncertainty for the $i$-th cluster, respectively. This procedure was applied 
to six different intervals of {$\Delta$(R.A.)cos(Dec.),
each containing an equal number of star clusters, except for the fifth bin, 
which included twice as many clusters. Figure~\ref{fig7} shows the resulting 
distribution, where the red segments represent the mean value $\langle Z 
\rangle$ and the shaded region indicates the intrinsic dispersion W for the
corresponding each interval ($\Delta$(R.A.)cos(Dec.) bins).

The $\langle Z \rangle$ values do not exhibit a monotonic trend with position; 
instead, they show an oscillatory variation along the East/West axis, 
indicating a complex vertical disk structure in the southern periphery of the 
LMC. while $\langle Z \rangle$ values would seem to oscillate across the
mid-plane, their dispersion would tend to 
increase toward the West (corresponding to decreasing values of
 $\Delta$(R.A.)cos(Dec.), i.e., toward to LMC region closer
to the SMC. This trend leads to a broadening of the
 $Z$ distribution, providing evidence of an increasingly perturbed geometry as 
one approaches the SMC. Taken together, these characteristics would seem compatible
 with a corrugated structure as predicted by the theory for the aftermath
of galaxy tidal interaction \citep{Laporte2018, Binney2024, FanXu2025}.
While the exact nature of these structures requires additional study, our results are 
consistent with the LMC-SMC interaction scenarios proposed by \citet{besla2012role} 
and \citet{Besla2016}, in which the LMC disk responses and corrugations 
induced by tidal perturbations can develop.

Tidal interactions in the southern regions of the LMC have been previously studied. For 
instance, \citet{choi2018smashing} used red clump stars to map the three-dimensional 
structure of the galaxy, reporting the first detection of a warp in the LMC. Likewise,
 \citet{massana2024magellanic} employed low-resolution {\it Gaia} DR3 spectra to analyze
the galaxy's peripheral substructures, finding that the region called LMC Hooks hosts a 
stellar population strongly perturbed by tidal forces associated to the LMC-SMC interaction.
 Our results provide independent evidence that the outer regions of the LMC exhibit structural
 perturbations linked to this interaction.

\subsection{Chemical analysis}

The agreement observed in Figure 6 between the age-metallicity distribution obtained in this work 
and the relations previously reported for the LMC indicates that clusters located in the outer disk 
($R >$ 5$\degr$) follow a chemical evolution consistent with that observed for other stellar 
populations of the main body of the galaxy. Furthermore, the better agreement with the 
bursting-type formation 
scenario relative to the closed-box model suggests that the chemical evolution of the LMC was 
dominated by relatively rapid episodes of chemical enrichment during the most recent stages of its 
evolution. However, none of the models considered fully reproduces the distribution obtained in 
this work. The close agreement between the AMR of \citet{Piatti2013} and our 
results would seem to support a scenario in which the gas present during the most recent stages of the 
galaxy's evolution was chemically well mixed on a global scale.

This AMR scenario is also consistent with the results of \citet{massana2024magellanic}, who suggest 
that the peripheral regions of the galaxy 
are dominated primarily by the LMC's own gas, with limited contribution from SMC material. 
In contrast, some works \citep{besla2012role,olsen2011population} propose that the last approach 
between both galaxies $\sim$200 Myr ago could have left remnants of SMC material, including
 gas, stars, or clusters, in the southern region. However, our results show no evidence 
of clusters with chemical properties incompatible with those typically observed in the LMC,
 suggesting no clear indications of clusters captured from the SMC in the analyzed regions. \\

The combination of structural and chemical outcomes is a primary result of this work.
 While the spatial distribution indicates the outer disk was perturbed by the interaction 
with the SMC, the chemical properties show these clusters continue to follow typical LMC 
relationships. This suggests the interaction was sufficient to dynamically deform the 
outer disk without necessarily transferring a significant population of chemically distinct
 clusters from the SMC. Overall, our results indicate the southern outer disk remains 
structurally and chemically part of the LMC, though it currently exhibits deformations 
compatible with recent tidal perturbations.

\begin{figure}
    \centering
    \includegraphics[width=\columnwidth]{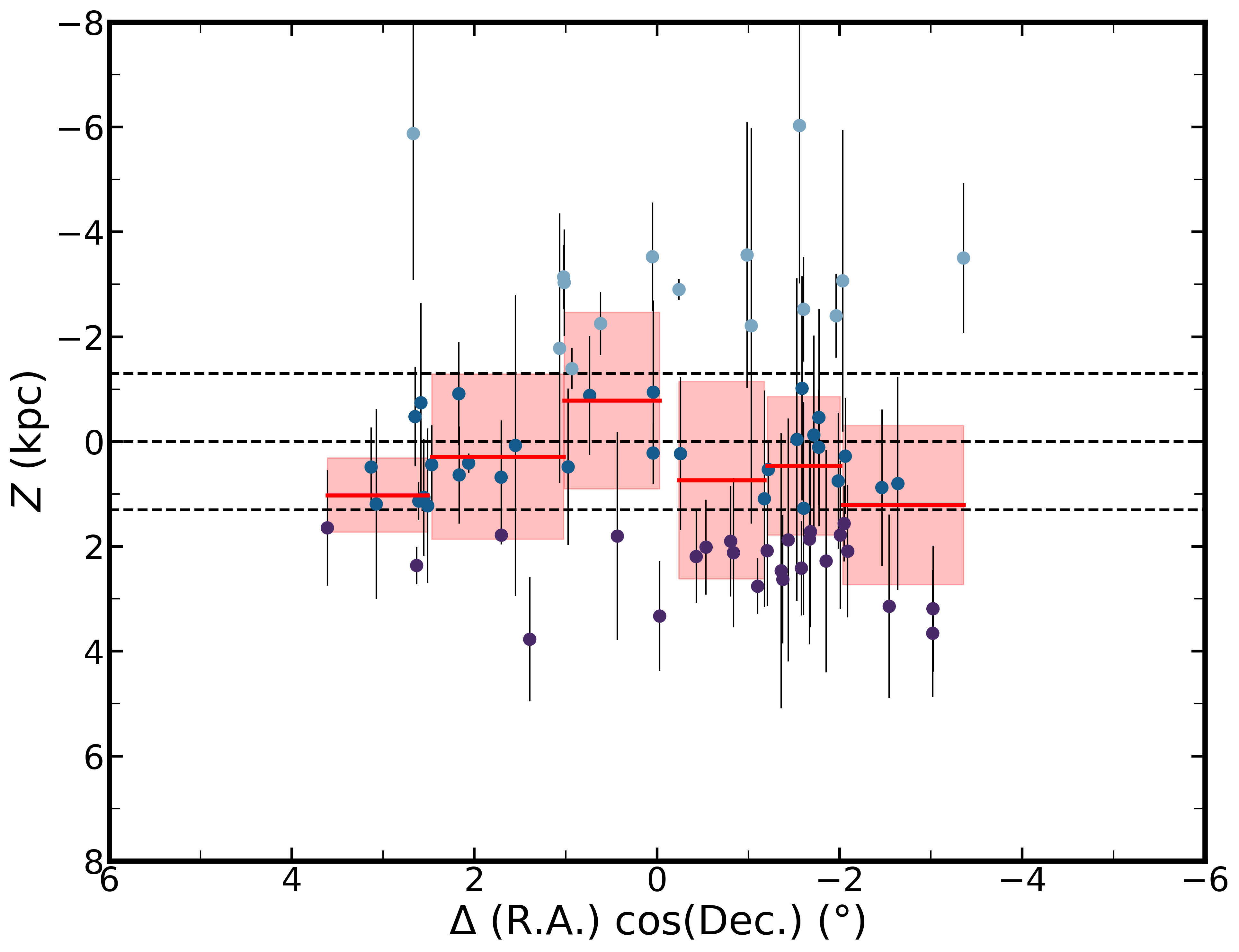}
    \caption{Edge-on view of the LMC disk, showing the vertical distribution of star clusters with 
    respect to the galaxy plane ($Z$). Blue, light-blue, and purple symbols denote clusters located 
    within the main body of the LMC disk, behind the LMC plane, and in front of the LMC plane, 
    respectively. Error bars correspond to uncertainties estimated through Monte Carlo simulations. The 
    red line indicates the mean height $\langle Z \rangle$, and the pink shaded region the intrinsic 
    dispersion $W$, derived from the maximum-likelihood analysis (see text for details). The dashed black 
    lines mark the LMC plane ($Z=0$) and the adopted disk thickness \citep{choi2018smashing}.}
    \label{fig7}
\end{figure}

\section{Summary and Conclusions}

In this work, we have performed a homogeneous analysis of 70 star clusters located in 
the southern outer disk of the LMC, covering a region beyond 
$5^{\circ}$ from the galactic center. Using photometric data 
from the SMASH DR2 survey, we derived precise ages, metallicities, and heliocentric 
distances for each cluster using the \texttt{ASTeCA} code. Our main conclusions are summarized 
as follows:\\
\begin{itemize}

\item The studied clusters possess ages ranging from 1.02 to 3.34 Gyr and metallicities
 [Fe/H] between -0.70 and -0.07 dex, with an average distance of 47.16 kpc.\\

\item We observed a clear spatial trend where clusters at greater heliocentric distances
 are predominantly located toward the western regions of the southern disk, which face 
the SMC.\\

\item The vertical distribution of heights ($Z$) relative to the LMC mid-plane reveals
 an oscillatory variation along the East/West axis. Furthermore, the intrinsic vertical 
dispersion ($W$) of the cluster population tends to increase toward the West, reflecting 
a broadening of the $Z$ distribution as the clusters approach the SMC.\\
 
\item These geometric features (oscillatory vertical variations and increased 
dispersion) provide independent evidence of a corrugated structure in the LMC southern outer
 disk. These findings are consistent with tidal interaction scenarios where the LMC
 disk develops vertical responses and corrugations induced by recent encounters with 
the SMC \citep{besla2012role, Besla2016, Laporte2018, Binney2024, FanXu2025}.\\

\item Chemically, the clusters follow the AMR typical 
of the LMC's own field and cluster populations. The observed distribution is 
well-reproduced by a bursting-type formation model \citep{pagel1998chemical}, suggesting that the
 gas in the outer disk was chemically well-mixed during recent enrichment episodes.\\

\item We found no evidence of clusters with chemical properties inconsistent with the
 LMC baseline, indicating that while tidal interactions were sufficient to dynamically 
deform the disk, they did not lead to the significant transfer of chemically distinct 
clusters from the SMC.\\
\end{itemize}

Overall, our results demonstrate that the southern outer disk of the LMC remains 
structurally and chemically part of the galaxy, though it currently exhibits substantial
 deformations and vertical oscillations consistent with recent tidal perturbations.

\section*{Data availability}
The CMD isochrone-fit figures for the clusters in the sample not shown in the main text (Sect. 4) are available at Zenodo via \href{https://doi.org/10.5281/zenodo.22776595}{https://doi.org/10.5281/zenodo.22776595}.

\begin{acknowledgements}
We thank the referee for the thorough reading of the manuscript and
timely suggestions to improve it.

Data for reproducing the figures and analysis in this work will be available upon request
to the first author.

\end{acknowledgements}

\begin{appendix}
\onecolumn

\section{Star cluster properties}

We here present the information gathered from the literature and the
values estimated in this work for astrophysical properties of the studied star clusters.

\begin{table}[h]
\caption{Astrophysical parameters of star clusters taken from the literature.}
\label{Table1}
\begin{tabular}{l c c c c}
\hline\hline
Star cluster & (m-M)$_{0}$ & log(age /year) & [Fe/H] & Ref.\\
        &  (mag)   &                & (dex)   &   \\
\hline
SL5, LW8, KMHK14                & 18.50$\pm$0.05& 9.55$\pm$0.05& -0.18$\pm$0.13& 1 \\
LW15, KMHK25                    & --            & --           & -0.42$\pm$0.20& 2 \\
SL13, LW17, KMHK31              & 18.60$\pm$0.04& 9.20$\pm$0.08& -0.10$\pm$0.11& 1 \\
SL28, LW47, ESO32-19, KMHK71    & --            & 9.18$\pm$0.09& -0.37$\pm$0.20& 3 \\
SL61, LW79, ESO32-20, KMHK178   & 18.53$\pm$0.06& 9.39$\pm$0.05& -0.44$\pm$0.19& 4 \\
OHSC3, KMHK362                  & 18.42$\pm$0.09& 9.25$\pm$0.06& -0.70$\pm$0.24& 4 \\
SL192, LW121, KMHK489           & --            & --           & -0.41$\pm$0.20& 2 \\
IC2134, SL437, LW198, ESO33-19  & --            & 9.00$\pm$0.13&       --      & 5 \\
SL451, LW206, KMHK883           & --            & 9.34$\pm$0.06& -0.70$\pm$0.20& 5 \\
SL455, LW207, KMHK891           & --            & 9.30$\pm$0.06& -0.77$\pm$0.20& 2 \\
LW231, KMHK1031                 & --            & 8.90$\pm$0.16& -0.70$\pm$0.20& 6 \\
IC2140, SL581, LW241, ESO33-24  & --            & 9.40$\pm$0.10& -0.84$\pm$0.22& 7 \\
IC2146, SL632, LW258, ESO33-26  & 18.42$\pm$0.15& 9.32$\pm$0.20& -0.40$\pm$0.15& 8 \\
LW263, KMHK1208                 & 18.50$\pm$0.05& 9.30$\pm$0.15& -0.10$\pm$0.11& 1 \\
NGC2161, SL789, LW337, ESO33-31 & 18.50$\pm$0.06& 9.20$\pm$0.20& -0.28$\pm$0.16& 1 \\
NGC2190, SL819, LW357, ESO33-36 & --            & 8.91$\pm$0.07& -0.12$\pm$0.20& 9 \\
SL828, LW367, ESO34-01, KMHK1616& --            & --           & -0.77$\pm$0.20& 2 \\
NGC2203, SL836, LW380, ESO34-04 & 18.38$\pm$0.15& 9.29$\pm$0.02& -0.30$\pm$0.15& 8 \\
SL835, LW379, KMHK1640          & --            & 9.00$\pm$0.17& -0.70$\pm$0.30& 10\\
NGC2209, SL849, LW408 ESO34-06  & 18.39$\pm$0.15& 9.16$\pm$0.02& -0.40$\pm$0.20& 8 \\
OHSC33, KMHK1714                & --            & 9.14$\pm$0.09& -1.00$\pm$0.20& 5 \\
\hline
\end{tabular}

\noindent References: 
(1) \citet{perren2017},
(2) \citet{olszewski1991spectroscopy}, 
(3) \citet{geisler1997search}, 
(4) \citet{maia2019viscacha}, 
(5) \citet{bica1998}, 
(6) \citet{palma2011parametros}, 
(7) \citet{piatti2011first}, 
(8) \citet{milone2023hubble}, 
(9) \citet{girardi1995age}, 
(10) \citet{livanou2013age}.

\end{table}

\begin{table}[h]
\caption{Derived astrophysical parameters of star clusters in this work.}
\label{Table2}
\begin{tabular}{l c c c c c}      
\hline\hline               
Star cluster & A$_V$$^{a}$ & (m-M)$_{0}$ & d    & log(age /year) & [Fe/H] \\
        &          &   (mag)       &  (kpc) &               & (dex)    \\
\hline                     
               SL5, LW8, KMHK14  &0.11&18.63$\pm$0.07&53.21$\pm$1.71&9.31$\pm$0.04&-0.34$\pm$0.07  \\
                   LW15, KMHK25  &0.09&18.26$\pm$0.07&44.87$\pm$1.45&9.30$\pm$0.03&-0.42$\pm$0.02  \\
              SL13, LW17, KMHK31 &0.09&18.24$\pm$0.07&44.46$\pm$1.43&9.27$\pm$0.04&-0.25$\pm$0.06  \\
    SL28, LW47, ESO32-19, KMHK71 &0.10&18.39$\pm$0.11&47.64$\pm$2.41&9.21$\pm$0.04&-0.31$\pm$0.03  \\
              SL29, LW50, KMHK81 &0.12&18.37$\pm$0.08&47.21$\pm$1.74&9.34$\pm$0.03&-0.56$\pm$0.07  \\
                   LW62, KMHK96  &0.11&18.26$\pm$0.10&44.87$\pm$2.07&9.24$\pm$0.02&-0.25$\pm$0.08  \\
             SL53, LW73, KMHK160 &0.11&18.29$\pm$0.07&45.50$\pm$1.47&9.35$\pm$0.01&-0.58$\pm$0.04  \\
   SL61, LW79, ESO32-20, KMHK178 &0.12&18.32$\pm$0.04&46.13$\pm$0.85&9.30$\pm$0.02&-0.47$\pm$0.07  \\
   SL74, LW82, ESO32-21, KMHK210 &0.11&18.40$\pm$0.01&47.86$\pm$0.22&9.24$\pm$0.01&-0.48$\pm$0.01  \\
             SL80, LW88, KMHK220 &0.11&18.57$\pm$0.14&51.76$\pm$3.34&9.09$\pm$0.05&-0.38$\pm$0.10  \\
                 OHSC1, KMHK231  &0.14&18.53$\pm$0.04&50.82$\pm$0.94&9.52$\pm$0.01&-0.40$\pm$0.01  \\
SL84, LW89, ESO32-22, KMHK232    &0.12&18.37$\pm$0.07&47.21$\pm$1.52&9.11$\pm$0.02&-0.43$\pm$0.07  \\
          H88-15, OHSC2, KMHK238 &0.09&18.32$\pm$0.08&46.13$\pm$1.70&9.06$\pm$0.05&-0.31$\pm$0.08  \\
            SL118, LW94, KMHK323 &0.10&18.29$\pm$0.12&45.50$\pm$2.51&9.23$\pm$0.02&-0.33$\pm$0.11  \\
                       KMHK343   &0.10&18.43$\pm$0.11&48.53$\pm$2.46&9.11$\pm$0.04&-0.43$\pm$0.05  \\
                   LW95, KMHK344 &0.13&18.40$\pm$0.01&47.86$\pm$0.22&9.14$\pm$0.01&-0.48$\pm$0.01  \\
                 OHSC3, KMHK362  &0.11&18.41$\pm$0.10&48.08$\pm$2.21&9.08$\pm$0.02&-0.34$\pm$0.08  \\
                 LW106, KMHK399  &0.13&18.45$\pm$0.11&48.98$\pm$2.48&9.26$\pm$0.02&-0.55$\pm$0.04  \\
                 LW107, KMHK396  &0.11&18.31$\pm$0.11&45.92$\pm$2.33&9.18$\pm$0.03&-0.36$\pm$0.04  \\
               H88-65            &0.10&18.32$\pm$0.10&46.13$\pm$2.12&9.07$\pm$0.08&-0.26$\pm$0.09  \\
                 LW110, KMHK414  &0.10&18.27$\pm$0.05&45.08$\pm$1.04&9.20$\pm$0.07&-0.30$\pm$0.10  \\
                       KMHK417   &0.11&18.34$\pm$0.11&46.56$\pm$2.36&9.29$\pm$0.07&-0.27$\pm$0.08  \\
                 OHSC4, KMHK426  &0.08&18.70$\pm$0.14&54.95$\pm$3.54&9.29$\pm$0.01&-0.47$\pm$0.02  \\
           SL166, LW111, KMHK419 &0.09&18.54$\pm$0.01&51.05$\pm$0.24&9.34$\pm$0.01&-0.48$\pm$0.01  \\
                       KMHK453   &0.11&18.29$\pm$0.13&45.50$\pm$2.72&9.16$\pm$0.21&-0.40$\pm$0.17  \\
                  \hline          
\end{tabular}

\noindent (a): Data taken from SMASH. 
\end{table}
 
\clearpage
 
\setcounter{table}{1}
\begin{table}[h]
\caption{Continued.}
\begin{tabular}{l c c c c c}      
\hline\hline               
Star cluster & A$_V$$^{a}$ & (m-M)$_{0}$ & d    & log(age /year) & [Fe/H] \\
        &          &   (mag)       &  (kpc) &               & (dex)    \\
\hline

                       KMHK447   &0.11&18.41$\pm$0.16&48.08$\pm$3.54&9.26$\pm$0.04&-0.46$\pm$0.01  \\
                         OHSC5   &0.11&18.26$\pm$0.15&44.87$\pm$3.10&9.05$\pm$0.11&-0.36$\pm$0.19  \\
        SL192, LW121, KMHK489    &0.09&18.26$\pm$0.07&44.87$\pm$1.45&9.26$\pm$0.04&-0.41$\pm$0.03  \\
                         OHSC7   &0.10&18.33$\pm$0.11&46.34$\pm$2.35&9.08$\pm$0.02&-0.27$\pm$0.03  \\
                  OHSC6, KMHK532 &0.09&18.37$\pm$0.01&47.21$\pm$0.22&9.31$\pm$0.01&-0.44$\pm$0.01  \\
                  LW132, KMHK542 &0.09&18.29$\pm$0.01&45.50$\pm$0.21&9.22$\pm$0.01&-0.32$\pm$0.01  \\
                 LW137, KMHK558  &0.09&18.24$\pm$0.03&44.46$\pm$0.61&9.28$\pm$0.03&-0.45$\pm$0.05  \\
                 SL248, KMHK578  &0.10&18.50$\pm$0.19&50.12$\pm$4.39&9.35$\pm$0.07&-0.52$\pm$0.14  \\
                 LW147, KMHK604  &0.08&18.58$\pm$0.12&52.00$\pm$2.87&9.40$\pm$0.07&-0.44$\pm$0.11  \\
           SL295, LW153, KMHK639 &0.09&18.27$\pm$0.08&45.08$\pm$1.66&9.14$\pm$0.03&-0.38$\pm$0.05  \\
                 LW155, KMHK647  &0.09&18.28$\pm$0.06&45.29$\pm$1.25&9.10$\pm$0.02&-0.27$\pm$0.05  \\
                 OHSC8, KMHK708  &0.11&18.28$\pm$0.05&45.29$\pm$1.04&9.18$\pm$0.01&-0.25$\pm$0.01  \\
                 OHSC9, KMHK733  &0.09&18.25$\pm$0.05&44.67$\pm$1.03&9.35$\pm$0.03&-0.51$\pm$0.06  \\
                OHSC11, KMHK803  &0.11&18.37$\pm$0.08&47.21$\pm$1.74&9.40$\pm$0.04&-0.53$\pm$0.08  \\
           SL400, LW188, KMHK809 &0.08&18.52$\pm$0.01&50.58$\pm$0.23&9.30$\pm$0.01&-0.47$\pm$0.02  \\
  IC2134, SL437, LW198, ESO33-19 &0.09&18.19$\pm$0.06&43.45$\pm$1.20&9.09$\pm$0.02&-0.26$\pm$0.04  \\
           SL451, LW206, KMHK883 &0.08&18.42$\pm$0.09&48.31$\pm$2.00&9.38$\pm$0.06&-0.49$\pm$0.10  \\
           SL455, LW207, KMHK891 &0.09&18.35$\pm$0.03&46.77$\pm$0.64&9.26$\pm$0.01&-0.47$\pm$0.04  \\
                OHSC16, KMHK881  &0.10&18.56$\pm$0.05&51.52$\pm$1.19&9.35$\pm$0.02&-0.47$\pm$0.03  \\
                LW231, KMHK1031  &0.08&18.27$\pm$0.11&45.08$\pm$2.28&9.09$\pm$0.05&-0.38$\pm$0.12  \\
  IC2140, SL581, LW241, ESO33-24 &0.07&18.48$\pm$0.01&49.66$\pm$0.23&9.44$\pm$0.01&-0.70$\pm$0.01  \\
          SL603, LW249, KMHK1151 &0.09&18.40$\pm$0.01&47.86$\pm$0.22&9.33$\pm$0.03&-0.68$\pm$0.07  \\
  IC2146, SL632, LW258, ESO33-26 &0.08&18.44$\pm$0.02&48.75$\pm$0.45&9.35$\pm$0.01&-0.53$\pm$0.04  \\
                LW263, KMHK1208  &0.12&18.52$\pm$0.05&50.58$\pm$1.16&9.26$\pm$0.01&-0.07$\pm$0.02  \\
IC2148, SL642, LW265, ESO33-28   &0.06&18.33$\pm$0.08&46.34$\pm$1.71&9.11$\pm$0.02&-0.41$\pm$0.06  \\
          SL647, LW267, KMHK1223 &0.06&18.52$\pm$0.03&50.58$\pm$0.70&9.26$\pm$0.02&-0.27$\pm$0.08  \\
                LW270, KMHK1243  &0.08&18.45$\pm$0.13&48.98$\pm$2.93&9.20$\pm$0.08&-0.61$\pm$0.16  \\
          SL703, LW290, KMHK1352 &0.11&18.15$\pm$0.07&42.66$\pm$1.38&9.12$\pm$0.00&-0.43$\pm$0.02  \\
          SL737, LW312, KMHK1432 &0.08&18.34$\pm$0.15&46.56$\pm$3.22&9.36$\pm$0.04&-0.38$\pm$0.10  \\
                LW315, KMHK1446  &0.12&18.32$\pm$0.06&46.13$\pm$1.27&9.14$\pm$0.02&-0.41$\pm$0.02  \\
          SL754, LW322, KMHK1473 &0.09&18.25$\pm$0.01&44.67$\pm$0.21&9.29$\pm$0.01&-0.58$\pm$0.01  \\
          SL783, LW333, KMHK1535 &0.14&18.33$\pm$0.01&46.34$\pm$0.21&9.25$\pm$0.01&-0.36$\pm$0.01  \\
 NGC2161, SL789, LW337, ESO33-31 &0.13&18.32$\pm$0.05&46.13$\pm$1.06&9.15$\pm$0.01&-0.35$\pm$0.09  \\
  IC2161, SL802, LW345, ESO33-35 &0.10&18.39$\pm$0.05&47.64$\pm$1.10&9.29$\pm$0.06&-0.47$\pm$0.09  \\
 NGC2190, SL819, LW357, ESO33-36 &0.13&18.32$\pm$0.04&46.13$\pm$0.85&9.11$\pm$0.01&-0.42$\pm$0.03  \\
               OHSC29, KMHK1604  &0.14&18.28$\pm$0.08&45.29$\pm$1.67&9.01$\pm$0.05&-0.07$\pm$0.07  \\
                LW360, KMHK1608  &0.15&18.29$\pm$0.06&45.50$\pm$1.26&9.06$\pm$0.07&-0.40$\pm$0.14  \\
SL828, LW367, ESO34-01, KMHK1616 &0.14&18.22$\pm$0.02&44.06$\pm$0.41&9.05$\pm$0.02&-0.30$\pm$0.05  \\
               OHSC30, KMHK1620  &0.11&18.64$\pm$0.13&53.46$\pm$3.20&9.36$\pm$0.02&-0.60$\pm$0.01  \\
                LW373, KMHK1625  &0.12&18.28$\pm$0.02&45.29$\pm$0.42&9.15$\pm$0.01&-0.39$\pm$0.01  \\
 NGC2203, SL836, LW380, ESO34-04 &0.09&18.37$\pm$0.10&47.20$\pm$2.17&9.30$\pm$0.00&-0.43$\pm$0.01  \\
          SL835, LW379, KMHK1640 &0.10&18.36$\pm$0.05&46.99$\pm$1.08&9.06$\pm$0.04&-0.27$\pm$0.08  \\
          SL848, LW405, KMHK1664 &0.14&18.28$\pm$0.10&45.29$\pm$2.09&9.06$\pm$0.03&-0.20$\pm$0.08  \\
 NGC2209, SL849, LW408, ESO34-06 &0.12&18.32$\pm$0.04&46.13$\pm$0.84&9.15$\pm$0.01&-0.42$\pm$0.02  \\
               OHSC33, KMHK1714  &0.09&18.25$\pm$0.06&44.67$\pm$1.23&9.07$\pm$0.04&-0.27$\pm$0.08  \\
\hline          
\end{tabular}

\end{table}

\twocolumn

\end{appendix}

\end{document}